\documentstyle[prd,preprint,tighten,aps,eqsecnum,amssymb,newlfont,epsfig]{revtex}
\begin{document}
\draft
\preprint{
\begin{tabular}{r}
DFTT 50/98\\
hep-ph/9808405\\
August 25, 1998
\end{tabular}
}
\title{Statistical interpretations of the null result\\
of the KARMEN~2 experiment}
\author{C. Giunti}
\address{INFN, Sezione di Torino, and Dipartimento di Fisica Teorica,
Universit\`a di Torino,\\
Via P. Giuria 1, I--10125 Torino, Italy}
\maketitle
\begin{abstract}
Several possible statistical interpretations
of the null result of the KARMEN~2 neutrino oscillation experiment
are discussed with the aim of clarifying the implications
of the fact that KARMEN~2 did not observe any
of the expected background events.
The formalism that allows to take into account the
error of the expected mean background
in a Poisson process with background is presented
and applied to the statistical analysis of the KARMEN~2 null result.
The possibility of ignoring the expected mean background
calculated for the KARMEN~2 experiment
is discussed
and it is shown that the resulting exclusion curves are ultra-conservative,
but may be a safe choice in such a controversial case.
It is also shown that the sensitivity curves of neutrino oscillation experiments
cannot be considered as exclusion curves.
The implications of the different statistical analyses of the KARMEN~2 null result
for the
compatibility of the results of the KARMEN~2 and LSND experiments
is discussed.
\end{abstract}


\section{Introduction}
\label{Introduction}

The KARMEN collaboration has reported recently~\cite{KARMEN}
a null result of the KARMEN~2 experiment in the search for
neutrino oscillations~\cite{Pontecorvo,neutrino oscillations}
in the channel
$\bar\nu_\mu\to\bar\nu_e$.
This result is very interesting because
the KARMEN neutrino oscillation experiment
is sensitive to the same region in the plane of the neutrino mixing parameters
$\sin^22\theta$ and $\Delta{m}^2$
as the LSND experiment~\cite{LSND}
whose results provide an evidence in favor of
$\bar\nu_\mu\to\bar\nu_e$
and
$\nu_\mu\to\nu_e$
oscillations.
Hence,
the statistical interpretation of the
KARMEN~2 null result
(as well as that of the LSND result)
is crucial in order to obtain an indication on the
compatibility or incompatibility
of the different results of the two experiments.

The KARMEN experiment~\cite{Drexlin} is located at
at the ISIS spallation neutron facility of the Rutherford Laboratories (UK).
The proton beam produced at ISIS is pulsed in time:
two 100 ns wide proton pulses separated by 330 ns
are produced every 20 ms.
The proton pulses are directed on a target where they produce
positive pions
(the negative pions are absorbed in the source before decaying)
which decay at rest according to the
process
\begin{equation}
\setlength{\arraycolsep}{0pt}
\begin{array}{rll} \displaystyle
\pi^+ \to
\null & \null \displaystyle
\mu^+
\null & \null \displaystyle
+ \nu_\mu
\\ \displaystyle
\null & \null \displaystyle
\downarrow
\null & \null \displaystyle
\\ \displaystyle
\null & \null \displaystyle
e^+
\null & \null \displaystyle
+ \nu_e + \bar\nu_\mu
\,,
\end{array}
\label{decay}
\end{equation}
producing an equal number of
$\nu_\mu$, $\nu_e$ and $\bar\nu_\mu$.
The KARMEN experiment searches for
$\bar\nu_e$
produced by
$\bar\nu_\mu\to\bar\nu_e$
oscillations
at a mean distance of 17.6 m.
The time structure of the neutrino beam
is important for the identification of the neutrino induced reactions
in the KARMEN detector
and for the effective suppression of the cosmic ray background.
The KARMEN experiment started in 1990 and run until 1995 as KARMEN 1~\cite{Drexlin}.
In 1996 it was upgraded to KARMEN~2,
eliminating the main cosmic ray induced background component
in the search for
$\bar\nu_\mu\to\bar\nu_e$
oscillations~\cite{Armbruster}.

So far the KARMEN~2 experiment
measured no events~\cite{KARMEN},
with an expected background of $ 2.88 \pm 0.13 $ events.
This null result has been analyzed with the following statistical methods:

\begin{description}

\item[Bayesian Approach.]
This method is recommended by the Particle Data Group~\cite{PDG96,PDG98}
and has been used by the KARMEN collaboration~\cite{KARMEN}.
The resulting upper limit for the mean $\mu$ of true neutrino oscillation events
is 2.3 and the corresponding exclusion curve is reproduced in Fig.~\ref{fig1}
(the solid curve passing through the filled squares).

\item[Unified Approach.]
This method has been proposed recently by Feldman and Cousins~\cite{FC98}
and has already been adopted
by the Particle Data Group~\cite{PDG98}
as the new statistical standard
for the calculation of frequentist confidence intervals
and upper limits
\emph{with the correct coverage}.
The Unified Approach is very attractive because
it allows to construct
classical confidence belts which ``unify the treatment
of upper confidence limits for null results
and two-sided confidence intervals for non-null results''.
On the other hand,
the Unified Approach has the undesirable feature that
when the number of observed events is smaller than the expected background,
the upper limits for the mean $\mu$ of true neutrino oscillation events
decreases rapidly when the background increases.
Hence,
by observing less events than the expected background
an experiment can establish a very stringent upper bound on $\mu$
even if it is not sensitive to such small values of $\mu$.
This is what happens in the Unified Approach analysis of the KARMEN~2 null result:
the upper limit for the mean $\mu$ of true neutrino oscillation events
is 1.1 and the corresponding exclusion curve~\cite{KARMEN}
reproduced in Fig.~\ref{fig1}
(the solid curve passing through the filled circles)
is very stringent,
even if the KARMEN~2 experiment is not sensitive to the
excluded region of the neutrino oscillation parameters
(see Ref.~\cite{Armbruster}).

\item[New Ordering Approach.]
This recently proposed method~\cite{CG98}
is based on a new ordering principle for the construction
of the classical confidence belt
which has all the desirable properties
of the one calculated with the Unified Approach
and in addition minimizes the effect
on the resulting confidence intervals
of the observation of less background events than expected.
Hence, it is appropriate for the statistical interpretation of the
null result of the KARMEN~2 experiment.
The resulting upper limit for the mean $\mu$ of true neutrino oscillation events
in the KARMEN~2 experiment
is 1.9 and the corresponding exclusion curve~\cite{CG98}
is reproduced in Fig.~\ref{fig1}
(the solid curve passing through the filled triangles).

\end{description}

The Unified Approach and the New Ordering Approach are reviewed
in Section~\ref{Poisson processes with background}.
From Fig.~\ref{fig1}
one can see that the Bayesian Approach and the Unified Approach
yield rather different exclusion from the null result of the KARMEN~2 experiment.
This difference is crucial in order to assess the compatibility
of the KARMEN~2 null result with the positive result of the LSND experiment,
whose 90\% CL
allowed region~\cite{LSND} is shown as the shadowed area in Fig.~\ref{fig1}.
The bayesian exclusion curve of KARMEN~2
is compatible with a large part of the LSND-allowed region,
whereas the exclusion curve obtained with the Unified Approach
is incompatible with almost all the LSND-allowed region.
The exclusion curve obtained with the New Ordering Approach
lies close to the bayesian exclusion curve
and tends to support the compatibility of the
KARMEN~2 and LSND results.
This is a desirable achievement.
Furthermore,
it is important to emphasize that,
as explained in Section~\ref{Poisson processes with background},
\emph{the New Ordering Approach
gives a correct frequentist coverage as the Unified Approach}.

In view of the uncertainty of the physical meaning of the KARMEN~2 null result
induced by the significant difference between the exclusion curve
obtained in the Unified Approach on one hand
and the exclusion curves obtained with the Bayesian Approach and
with the New Ordering Approach on the other hand,
it is interesting to explore other possibilities
for the statistical interpretation of the KARMEN~2 null result.

In Sections~\ref{Unknown Background}--\ref{Sensitivity}
I discuss the following
alternative statistical interpretations of the null result of the KARMEN~2 experiment,
which take into account the problem of the non-observation
of expected background events in the KARMEN~2 experiment:

\begin{description}

\item[Larger Background Error.]
Since no background events have been measured in the KARMEN~2 experiment,
it is possible to doubt of the correctness of the calculated
error for the mean expected background,
$ b = 2.88 \pm 0.13 $.
Hence,
one can consider the possibility
that the error has been underestimated,
for example,
by an order of magnitude.
This approach is discussed in Section~\ref{Larger Background Error},
together with a presentation of the appropriate formalism.

\item[Unknown Background.]
An extreme attitude towards the doubting of
the correctness of the calculated mean expected background
$ b = 2.88 \pm 0.13 $
is to ignore it and to assume that the background is unknown.
In this case the
KARMEN~2 null measurement
gives an estimation of the mean $ \mu + b $
of signal plus background events.
This approach is discussed in Section~\ref{Unknown Background}.

\item[Sensitivity.]
The sensitivity of an experiment is
defined as ``the average upper limit that would be obtained
by an ensemble of experiments with the expected background
and no true signal''~\cite{FC98}.
The sensitivities of the KARMEN~2 experiment
calculated with the Unified Approach and with the New Ordering Approach
are shown in Fig.~\ref{fig1}
(the solid curves passing through the empty circles and triangles, respectively).
One could think of considering the sensitivity curve of the KARMEN~2
experiment as its exclusion curve.
This approach is discussed and
shown to be incorrect in Section~\ref{Sensitivity}.

\end{description}

\section{Poisson processes with background}
\label{Poisson processes with background}

The events measured in the KARMEN experiment follow a Poisson distribution.
The probability to observe a number $n$
of events in a Poisson process
consisting in signal events with mean $\mu$
and background events with known mean $b$
is
\begin{equation}
P(n|\mu;b)
=
\frac{1}{n!} \ (\mu+b)^n \, e^{-(\mu+b)}
\,.
\label{poisson}
\end{equation}
The classical frequentist method for obtaining the confidence interval
for the unknown parameter $\mu$
is based on Neyman's method to construct a \emph{confidence belt}.
This confidence belt is the region in the $n$--$\mu$ plane
lying between the two curves $n_1(\mu;b,\alpha)$ and $n_2(\mu;b,\alpha)$
such that for each value of $\mu$
\begin{equation}
P(n\in[n_1(\mu;b,\alpha),n_2(\mu;b,\alpha)]|\mu;b)
\equiv
\sum_{n=n_1(\mu;b,\alpha)}^{n_2(\mu;b,\alpha)} P(n|\mu;b)
=
\alpha
\,,
\label{CL}
\end{equation}
where
$\alpha$ is the desired confidence level.
The two curves
$n_1(\mu;b,\alpha)$ and $n_2(\mu;b,\alpha)$
are required to be monotonic functions of $\mu$
and can be inverted to yield the corresponding curves
$\mu_1(n;b,\alpha)$ and $\mu_2(n;b,\alpha)$.
Then,
if a number $n_{\mathrm{obs}}$ of events is measured,
the confidence interval for $\mu$ is
$[\mu_2(n_{\mathrm{obs}};b,\alpha),\mu_1(n_{\mathrm{obs}};b,\alpha)]$.
This method guarantees by construction the \emph{correct coverage},
\textit{i.e.}
the fact that the resulting confidence interval
$[\mu_2(n_{\mathrm{obs}};b,\alpha),\mu_1(n_{\mathrm{obs}};b,\alpha)]$
is a member of a set of confidence intervals
obtained with an ensemble of similar experiments
that
contain the true value of $\mu$ with a probability $\alpha$.
Actually,
in the case of a Poisson process,
since $n$ is an integer,
the relation (\ref{CL})
can only be approximately satisfied and in practice the chosen
\emph{acceptance intervals}
$[n_1(\mu;b,\alpha),n_2(\mu;b,\alpha)]$
are the smallest intervals such that
\begin{equation}
P(n\in[n_1(\mu;b,\alpha),n_2(\mu;b,\alpha)]|\mu;b)
\geq
\alpha
\,.
\label{CLP}
\end{equation}
This choice introduces an overcoverage for some values of $\mu$
and the resulting confidence intervals
are \emph{conservative}.
As emphasized in Ref.~\cite{FC98}
conservativeness is an undesirable but unavoidable property
of the confidence intervals in the case of a Poisson process
(it is undesirable because it
implies a loss of power
in restricting the allowed range for the parameter $\mu$).

The construction of Neyman's confidence belt
\emph{is not unique},
because in general there are many different couples of curves
$n_1(\mu;b,\alpha)$ and $n_2(\mu;b,\alpha)$
that satisfy the relation (\ref{CL}).
Hence,
an additional criterion is needed in order to
define uniquely the acceptance intervals
$[n_1(\mu;b,\alpha),n_2(\mu;b,\alpha)]$.
The two common choices are 
\begin{equation}
P(n<n_1(\mu;b,\alpha)|\mu;b)
=
P(n>n_2(\mu;b,\alpha)|\mu;b)
=
\frac{1-\alpha}{2}
\,,
\label{central}
\end{equation}
which leads to
\emph{central confidence intervals}
and
\begin{equation}
P(n<n_1(\mu;b,\alpha)|\mu;b)
=
1-\alpha
\,,
\label{upper}
\end{equation}
which leads to
\emph{upper confidence limits}.
Central confidence intervals are appropriate for the
statistical description of the results of experiments reporting a positive result,
\textit{i.e.} the measurement of a number of events
significantly larger than the expected background.
On the other hand,
upper confidence limits are appropriate for the
statistical description of the results of experiments reporting a negative result,
\textit{i.e.} the measurement of a number of events
compatible with the expected background.
However,
Feldman and Cousins~\cite{FC98}
noticed that switching from central confidence level
to upper confidence limits or vice-versa
on the basis of the experimental data
(``flip-flopping'')
leads to undercoverage for some values of $\mu$,
which is a serious flaw for a frequentist method.

Feldman and Cousins \cite{FC98}
proposed an ordering principle
for the construction of the acceptance intervals
that is
based on likelihood ratios
and produces an automatic transition
from central confidence intervals to upper limits
when the number of observed events in a Poisson process with background
is of the same order or less than the expected background,
guaranteeing the correct frequentist coverage for all values of $\mu$.
The acceptance interval for each value of $\mu$
is calculated assigning at each value of $n$ a rank
obtained from the relative size of the ratio
\begin{equation}
R_{\mathrm{UA}}(n|\mu;b)
=
\frac{ P(n|\mu;b) }{ P(n|\mu_{\mathrm{best}};b) }
\,,
\label{RUA}
\end{equation}
where $\mu_{\mathrm{best}}=\mu_{\mathrm{best}}(n;b)$
(for a fixed $b$)
is the non-negative value of $\mu$ that
maximizes the probability
$P(n|\mu;b)$:
\begin{equation}
\mu_{\mathrm{best}}(n;b)
=
\mathrm{max}[0,n-b]
\,.
\label{best}
\end{equation}
As emphasized in Ref.~\cite{FC98},
``$R$ is a ratio of two likelihoods:
the likelihood of obtaining $n$ given the actual mean $\mu$,
and the likelihood of obtaining $n$
given the best-fit physically allowed mean''.
For each fixed value of $\mu$,
the rank of each value of $n$
is assigned in order of decreasing value of the ratio $R_{\mathrm{UA}}(n|\mu;b)$:
the value of $n$ which has bigger $R_{\mathrm{UA}}(n|\mu;b)$ has rank one,
the value of $n$ among the remaining ones
which has bigger $R_{\mathrm{UA}}(n|\mu;b)$ has rank two
and so on.
The acceptance interval for each value of $\mu$
is calculated by adding the values of $n$ in increasing order of rank
until the condition (\ref{CLP}) is satisfied.

The automatic transition from two-sided confidence intervals
to upper confidence limits for $ n \lesssim b $
is guaranteed in the Unified Approach by the fact that
$\mu_{\mathrm{best}}$
is always non-negative.
Indeed,
since
$ \mu_{\mathrm{best}}(n{\leq}b;b) = 0 $,
the rank of
$n \leq b$ for $\mu=0$
is one,
implying that the interval
$ 0 \leq n \leq b $
for $\mu=0$
is guaranteed to lie in the confidence belt.

Although the Unified Approach
solves brilliantly the problem of obtaining a
transition with correct coverage from two-sided confidence intervals
to upper confidence limits for $ n \lesssim b $,
it has the undesirable feature that when
$ n \lesssim b $
the upper bound 
$\mu_1(n;b,\alpha)$
decreases rather rapidly when $b$ increases
and stabilizes around a value close to 0.8
for large values of $b$.
From a physical point of view
this is rather disturbing,
because
\emph{a stringent upper bound for $\mu$
obtained by an experiment which has observed a number of events
significantly smaller than the expected background
is not due to the fact that the experiment is very sensitive to small values of $\mu$,
but to the fact that less background events than expected have been observed}.

This is the case of the null result of the KARMEN~2 experiment,
from which the Unified Approach yields an upper 90\% CL
confidence limit for the mean $\mu$ of neutrino oscillation events
of 1.1 events.
The corresponding exclusion curve in the plane of the neutrino mixing parameters
$\sin^22\theta$ and $\Delta{m}^2$
is shown in Fig.~\ref{fig1}
(the solid curve passing through the filled circles)
and one can see that it is significantly more stringent than the exclusion curve
obtained with the Bayesian Approach
(the solid curve passing through the filled squares).
The strictness of the Unified Approach exclusion curve
is due to the non-observation of the expected background events
and not to the sensitivity of the experiment
(see the discussion in Ref.~\cite{CG98}
and the sensitivity of the KARMEN experiment presented in Ref.\cite{Armbruster}).
This is clearly an undesirable result from a physical point of view,
because
\emph{the statistical interpretation of the data
produces an exaggeratedly stringent result that could lead to
incorrect physical conclusions}.

The discrepancy between the Bayesian Approach exclusion curve
and the Unified Approach exclusion curve
is worrying for a physicist,
because the Bayesian Approach exclusion curve
is compatible with a large part of the
LSND-allowed region
(the shadowed area in Fig.~\ref{fig1}),
whereas the Unified Approach exclusion curve
excludes almost all the LSND allowed region.

In Ref.~\cite{CG98} I have proposed
\emph{a new ordering principle}
for the construction of a classical confidence belt
which has all the desirable features of the one
in the Unified Approach
(\textit{i.e.}
an automatic transition with the correct coverage
from two-sided confidence intervals to
upper confidence limits when the observed number of events
is of the order or less than the expected background)
and in addition minimizes
the decrease of the upper confidence limit for a given $n$
as the mean expected background $b$ increases.
The new ordering principle is implemented as the
Feldman and Cousins ordering principle in the Unified Approach,
but for each value of $\mu$
the rank of each value of $n$
is calculated from the relative size of the likelihood ratio
\begin{equation}
R_{\mathrm{NO}}(n|\mu;b)
=
\frac{ P(n|\mu;b) }{ P(n|\mu_{\mathrm{ref}};b) }
\,,
\label{RNO}
\end{equation}
where the reference value $\mu_{\mathrm{ref}}=\mu_{\mathrm{ref}}(n;b)$
is taken to be the bayesian expected value for $\mu$:
\begin{equation}
\mu_{\mathrm{ref}}(n;b)
=
\int_0^\infty \mu \, P(\mu|n;b) \, \mathrm{d}\mu
=
n + 1
-
\left( \displaystyle \sum_{k=0}^{n} \frac{k\,b^k}{k!} \right)
\left( \displaystyle \sum_{k=0}^{n} \frac{b^k}{k!} \right)^{-1}
\,.
\label{mu_ref}
\end{equation}
Here $P(\mu|n;b)$
is the bayesian probability distribution for $\mu$
calculated assuming a constant prior for $\mu\geq0$
(see, for example, \cite{D'Agostini}):
\begin{equation}
P(\mu|n;b)
=
( b + \mu )^n \, e^{-\mu}
\left( \displaystyle n! \, \sum_{k=0}^{n} \frac{b^k}{k!} \right)^{-1}
\,.
\label{bayes}
\end{equation}
The obvious inequality
$
\sum_{k=0}^{n} k\,b^k/k!
\leq
n \sum_{k=0}^{n} b^k/k!
$
implies that
$ \mu_{\mathrm{ref}}(n;b) \geq 1 $.
Therefore,
$\mu_{\mathrm{ref}}(n;b)$
represents a reference value for $\mu$ that not only is non-negative,
as desired in order to have an automatic
transition from
two-sided intervals to upper limits,
but is even bigger or equal than one.
This is a desirable characteristics in order to
obtain a weak decrease of the upper confidence limit
for a given $n$
when the expected background $b$ increases.
Indeed,
it has been shown in Ref.~\cite{CG98}
that for
$ n \lesssim b $
the upper bound 
$\mu_1(n;b,\alpha)$
decreases rather weakly when $b$ increases
and stabilizes around a value close to 1.7
for large values of $b$.
This behaviour of $\mu_1(n;b,\alpha)$
is more acceptable for the physical interpretation of
experimental results
than the behaviour of $\mu_1(n;b,\alpha)$
in the Unified Approach.

The 90\% CL
upper confidence limit for the mean $\mu$ of neutrino oscillation events
following from the analysis of the KARMEN~2 null result
with the New Ordering Approach
is of 1.9 events
and the corresponding exclusion curve in the plane of the
neutrino mixing parameters $\sin^22\theta$ and $\Delta{m}^2$
is shown in Fig.~\ref{fig1}
(the solid curve passing through the filled triangles).
One can see that this exclusion curve,
although obtained with a method that guarantees
the correct frequentist coverage as the Unified Approach,
is significantly less stringent than the one obtained with the
Unified Approach
and lies close to the bayesian exclusion curve.

Hence,
the New Ordering Approach has solved the apparent conflict
between the frequentist and bayesian statistical
interpretation of the null result of the KARMEN~2 experiment:
\emph{by choosing an appropriate ordering principle
in the construction of the confidence belt,
the exclusion curve obtained with the frequentist method
is in reasonable agreement with the one obtained with the Bayesian Approach}.

Nevertheless,
some concern still remain on the interpretation of the fact that
the KARMEN~2 experiment
did not observe any of the expected background events.
Such concern will increase in the future if
the KAREN 2 experiment will continue to observe
significantly less events than
those expected
from the background. 
Hence,
it is interesting to explore other possibilities
for the statistical interpretation of the result
of the KARMEN2 experiment.
This is the aim of the following three Sections.

\section{Larger Background Error}
\label{Larger Background Error}

Since no background events have been measured in the KARMEN~2 experiment
and the mean expected background is
$ b = 2.88 \pm 0.13 $, 
it is possible to doubt of the correctness of the calculated
error for the mean expected background.
In this Section, I will
investigate which would be the physical implications
of the KARMEN~2 null result if the calculated error for
the mean expected background has been underestimated by an order of magnitude,
\textit{i.e.}
I will take $ b = 2.88 \pm 1.3 $.
I will assume a normal probability distribution function for the
mean expected background $b$:
\begin{equation}
f(b;\overline{b},\sigma_b)
=
\frac{ 1 }{ \sqrt{2\pi} \ \sigma_b }
\
\exp\left[ - \frac{ (b-\overline{b})^2 }{ 2 \, \sigma_b^2 } \right]
\,,
\label{normal}
\end{equation}
with
$\overline{b}=2.88$
and
$\sigma_b=1.3$.
For simplicity,
since
$ \sigma_b \lesssim \overline{b} / 3 $,
I will consider
$b$ varying from $-\infty$ to $+\infty$,
neglecting the small error introduced by considering
negative values of $b$.
This approximation allows an analytic solution of the integrals
involved in the calculation\footnote{If $b$ is restricted
to the interval $[0,+\infty)$
the exact normalization factor of
$f(b;\overline{b},\sigma_b)$
replacing $ 1 / \sqrt{2\pi} \sigma_b $
in Eq.(\ref{normal}) is $1/N(\overline{b},\sigma_b)$ with
\begin{displaymath}
N(\overline{b},\sigma_b)
=
\sqrt{ \frac{ \pi }{ 2 } }
\left[
1 + \mathrm{erf}\!\left( \frac{ \overline{b} }{ \sqrt{2} \ \sigma_b } \right)
\right]
\sigma_b
\,.
\end{displaymath}
In this case
$N(\overline{b},\sigma_b)$
and the integral in Eq.(\ref{pnmu1})
must be calculated numerically.}.

If $\mu$ is the mean of true signal events,
the probability $P(n|\mu;\overline{b},\sigma_b)$
to observe $n$ events is given by
\begin{equation}
P(n|\mu;\overline{b},\sigma_b)
=
\int
P(n|\mu;b)
\
f(b;\overline{b},\sigma_b)
\
\mathrm{d}b
\,,
\label{pnmu1}
\end{equation}
with the Poisson probability
$P(n|\mu;b)$
given in Eq.(\ref{poisson}).
Hence,
$P(n|\mu;\overline{b},\sigma_b)$
can be written as
\begin{equation}
P(n|\mu;\overline{b},\sigma_b)
=
\frac{1}{n!}
\
e^{-\mu}
\
\sum_{k=0}^{n}
\left( \begin{array}{c} n \\ k \end{array} \right)
\
\mu^{n-k}
\
I_k(\overline{b},\sigma_b)
\,,
\label{pnmu2}
\end{equation}
with
\begin{equation}
I_k(\overline{b},\sigma_b)
\equiv
\frac{ 1 }{ \sqrt{2\pi} \ \sigma_b }
\int_{-\infty}^{+\infty}
b^k
\
\exp\left[ - b - \frac{ (b-\overline{b})^2 }{ 2 \, \sigma_b^2 } \right]
\
\mathrm{d}b
\,.
\label{Ik1}
\end{equation}

The integral
$I_k(\overline{b},\sigma_b)$
can be written as
\begin{equation}
I_k(\overline{b},\sigma_b)
=
\exp\left[ - \overline{b} + \frac{\sigma_b^2}{2} \right]
\
\sum_{j=0}^{k}
\left( \begin{array}{c} k \\ j \end{array} \right)
( \overline{b} - \sigma_b^2 )^{k-j}
\
\sigma_b^j
\
m_j
\,,
\label{Ik2}
\end{equation}
where $m_j$
is the $j^{\mathrm{th}}$
central moment of the normal distribution with unit variance,
\begin{equation}
m_j
\equiv
\frac{ 1 }{ \sqrt{2\pi} }
\int_{-\infty}^{+\infty}
b^j
\
e^{ - b^2 / 2 }
\
\mathrm{d}b
\,.
\label{mom1}
\end{equation}
Taking into account that
$
\int b \, e^{ - b^2 / 2 } \, \mathrm{d}b
=
- e^{ - b^2 / 2 }
$,
the integral in Eq.(\ref{mom1})
can be calculated by parts,
yielding
\begin{equation}
m_j
=
\frac{ j! }{ (j/2)! \ 2^{j/2} }
\label{mom2}
\end{equation}
for $j$ even and
$ m_j = 0 $
for $j$ odd.

From Eqs.(\ref{pnmu2}), (\ref{Ik2}) and (\ref{mom2}),
for the probability of $n$ events one obtains
\begin{equation}
P(n|\mu;\overline{b},\sigma_b)
=
\exp\left[ - ( \mu + \overline{b} ) + \frac{\sigma_b^2}{2} \right]
\
\sum_{k=0}^{n}
\frac{ \mu^{n-k} }{ (n-k)! }
\
J_k(\overline{b},\sigma_b)
\,,
\label{pnm3}
\end{equation}
with
\begin{equation}
J_k(\overline{b},\sigma_b)
\equiv
\sum_{j=0}^{k/2}
\frac
{ ( \overline{b} - \sigma_b^2 )^{k-2j} \, \sigma_b^{2j} }
{ (k-2j)! \, j! \, 2^j }
\,.
\label{jeik}
\end{equation}

Equation (\ref{pnm3}) gives the formula for the
probability
$P(n|\mu;\overline{b},\sigma_b)$
to observe a number $n$
of events in a Poisson process
consisting in signal events with mean $\mu$
and background events with known mean $b=\overline{b}\pm\sigma_b$,
\textit{i.e.}
it replaces Eq.(\ref{poisson})
if the error $\sigma_b$ of the calculated mean background is not negligible
(but $ \sigma_b \lesssim \overline{b} / 3 $).
The construction of the confidence belt
follows the same procedure described in
Section~\ref{Poisson processes with background}
and now the confidence interval for $\mu$
corresponding to
a number $n_{\mathrm{obs}}$ of observed events is
$
[
\mu_2(n_{\mathrm{obs}};\overline{b},\sigma_b,\alpha)
,
\mu_1(n_{\mathrm{obs}};\overline{b},\sigma_b,\alpha)
]
$,
\textit{i.e.}
it depends on $\overline{b}$ and $\sigma_b$.
The acceptance intervals can be constructed following the same
principles discussed in
Section~\ref{Poisson processes with background},
\textit{i.e.}
one can construct the confidence belt for central confidence intervals
or upper confidence limits,
or the confidence belt in the Unified Approach
or in the New Ordering Approach.
In the remaining part of this Section I will present the formalism
for the implementation of 
the Unified Approach
and
of the New Ordering Approach
and I will discuss the corresponding implications for the statistical
interpretation of the KARMEN~2 null result.

The quantity $\mu_{\mathrm{best}}(n;\overline{b},\sigma_b)$
in the Unified Approach
is the value of $\mu$ that maximizes $P(n|\mu;\overline{b},\sigma_b)$
and
the acceptance interval for each value of $\mu$
is calculated assigning at each value of $n$ a rank
obtained from the relative size of the ratio
\begin{equation}
R_{\mathrm{UA}}(n|\mu;\overline{b},\sigma_b)
=
\frac{ P(n|\mu;\overline{b},\sigma_b) }{ P(n|\mu_{\mathrm{best}};\overline{b},\sigma_b) }
\,.
\label{RUA1}
\end{equation}
The value of
$\mu_{\mathrm{best}}(n;\overline{b},\sigma_b)$
can be easily calculated by hand
for $n=0,1,2$,
whereas
for higher values of $n$
it can be calculated numerically.
The resulting confidence belt for
$\overline{b}=2.88$
and
$\sigma_b=1.3$
is plotted in Fig.~\ref{fig2}
(the region between the two dashed lines),
where it is compared with the confidence belt for
$\sigma_b=0$
(the region between the two solid lines),
which practically coincides with the confidence belt for
$\sigma_b=0.13$.
The confidence belt for $\sigma_b=1.3$
is larger than the one for $\sigma_b=0$
because the integral in Eq.(\ref{pnmu1})
has the effect of flattening the probability
$P(n|\mu;\overline{b},\sigma_b)$
as a function of $n$
for fixed $\mu$
with respect to
$P(n|\mu;b=\overline{b})$
and
this flattening effect increases with the size of $\sigma_b$. 

From Fig.~\ref{fig2}
one can see that the upper limit $\mu_1$
for $n=0$
is of 1.2 events,
only slightly larger than the upper limit
of 1.0 events
obtained for $\sigma_b=0$.
Therefore,
the exclusion curve that follow from the null result
of the KARMEN~2 experiment
if $\sigma_b=1.3$
lies close to the
solid curve passing through the filled circles in Fig.~\ref{fig1}
(corresponding to $\mu_1=1.1$)\footnote{In the Unified Approach of Ref.~\cite{FC98}
$\sigma_b=0$
but the upper limit $\mu_1$
for $n=0$
is of 1.1 events because
$\mu_1$
is forced to be a non-increasing function of $b$.}.

In the New Ordering Approach
the acceptance interval for each value of $\mu$
is calculated assigning at each value of $n$ a rank
obtained from the relative size of the ratio
\begin{equation}
R_{\mathrm{NO}}(n|\mu;\overline{b},\sigma_b)
=
\frac{ P(n|\mu;\overline{b},\sigma_b) }{ P(n|\mu_{\mathrm{ref}};\overline{b},\sigma_b) }
\,,
\label{RNO1}
\end{equation}
where
the reference value
$\mu_{\mathrm{ref}}=\mu_{\mathrm{ref}}(n;\overline{b},\sigma_b)$
is the bayesian expected value for $\mu$.
The value of $\mu_{\mathrm{ref}}(n;\overline{b},\sigma_b)$
can be calculated analytically.
For the bayesian probability distribution function for $\mu$
with a constant prior,
\begin{equation}
P(\mu|n;\overline{b},\sigma_b)
=
\frac
{ P(n|\mu;\overline{b},\sigma_b) }
{ \int_0^\infty P(n|\mu;\overline{b},\sigma_b) \ \mathrm{d}\mu }
\,,
\label{bayes1}
\end{equation}
one obtains
\begin{equation}
P(\mu|n;\overline{b},\sigma_b)
=
e^{-\mu}
\left(
\sum_{k=0}^{n}
\frac{ \mu^{n-k} }{ (n-k)! }
\
J_k(\overline{b},\sigma_b)
\right)
\left(
\sum_{k=0}^{n}
J_k(\overline{b},\sigma_b)
\right)^{-1}
\,.
\label{bayes2}
\end{equation}
Hence,
the reference value $\mu_{\mathrm{ref}}(n;\overline{b},\sigma_b)$,
which is the bayesian expected value for $\mu$,
is given by
\begin{equation}
\mu_{\mathrm{ref}}(n;\overline{b},\sigma_b)
=
n + 1
-
\left(
\sum_{k=0}^{n}
k
\
J_k(\overline{b},\sigma_b)
\right)
\left(
\sum_{k=0}^{n}
J_k(\overline{b},\sigma_b)
\right)^{-1}
\,.
\label{mu_ref1}
\end{equation}
The resulting confidence belt in the New Ordering Approach for
$\overline{b}=2.88$ and $\sigma_b=1.3$
is shown in Fig.~\ref{fig3}
(the region between the two dashed lines),
where it is compared with the confidence belt for
$\sigma_b=0$
(the region between the two solid lines),
which practically coincides with the confidence belt for
$\sigma_b=0.13$.
One can see that the upper limit for $\mu$
if $n_{\mathrm{obs}}=0$
is of 2.0 events,
only slightly larger than the upper limit
of 1.9 events
obtained for $\sigma_b=0$.
Therefore,
the KARMEN~2 exclusion curve
for $\sigma_b=1.3$
lies close to the
solid curve passing through the filled triangles in Fig.~\ref{fig1}
(for $\sigma_b=1.3$
the ordinate of the exclusion curve
for high $\Delta{m}^2$'s
is
$ \sin^22\theta = 2.5 \times 10^{-3} $,
compared with
$ \sin^22\theta = 2.3 \times 10^{-3} $
for
$\sigma_b=0$).

The conclusion that one can reach from the results presented in this Section
is that the error of the calculated mean background
has little effect on the exclusion curves obtained
in the Unified Approach
and
in the New Ordering Approach
(at least for $ \sigma_b \lesssim \overline{b} / 3 $). 
Even if the error $\sigma_b$ of the mean background
in the KARMEN~2 experiment
is increased by an order of magnitude
with respect to the calculated one
($\sigma_b=0.13$),
the exclusion curves
in the Unified Approach
and
in the New Ordering Approach
are practically
equivalent to the corresponding curves calculated for $\sigma_b=0$.

\section{Unknown Background}
\label{Unknown Background}

Since the KARMEN~2 experiment did not observe any event,
with a mean expected background
$ b = 2.88 \pm 0.13 $,
and even increasing by an order of magnitude the estimated error
of the mean expected background does not help in solving the discrepancy
among the exclusion curves calculated with different statistical methods,
one can consider the extreme possibility of doubting
of the correctness of the calculated mean expected background
and ignore it.
In other words,
one can assume that the background is unknown
and that the KARMEN~2 null measurement
gives an estimation of the mean $ \mu + b $
of signal plus background events.

This approach leads to an upper confidence limit $ (\mu + b)_1 $
which depends only on the observed number of events.
Since $ b \geq 0 $,
the upper confidence limit $ (\mu + b)_1 $
for $ \mu + b $
is also a conservative upper confidence limit for $\mu$.
Although the derivation of a conservative upper confidence limit
is generally an undesirable result,
in controversial cases as that of the KARMEN~2 experiment it may be a safe
choice.

Actually,
in the implementation of the Unknown Background approach one has also
to chose the method for building the confidence belt.
I will consider the following four possibilities,
whose confidence belts are plotted in Fig.~\ref{fig4}:
\begin{description}

\item[Standard confidence belt for upper confidence limits.]
The resulting 90\% CL
confidence belt
is the region below the solid line in Fig.~\ref{fig4}.
This method implies an upper 90\% CL
confidence limit of 2.3 events for
$ \mu + b $ if the number of observed events $n_{\mathrm{obs}}$ is zero.
Taking it as a conservative upper confidence limit for $\mu$
leads to an exclusion curve coinciding with the bayesian exclusion curve
shown in Fig.~\ref{fig1}.
This exclusion curve is plotted in Fig.~\ref{fig5}
(the solid curve passing through the filled squares).

\item[Standard confidence belt for central confidence intervals.]
The resulting 90\% CL
confidence belt
is the region between the two dashed lines in Fig.~\ref{fig4}
and the 90\% CL
upper limit for
$ \mu + b $
is of 3.0 events if $n_{\mathrm{obs}}=0$.
The corresponding exclusion curve is plotted in Fig.~\ref{fig5}
(the solid curve passing through the dotted squares).

\item[Unified Approach.]
The 90\% CL
confidence belt
is the region between the two dash-dotted lines in Fig.~\ref{fig4}
and the upper limit for
$ \mu + b $
is of 2.4 events for $n_{\mathrm{obs}}=0$.
The corresponding exclusion curve is plotted in Fig.~\ref{fig5}
(the solid curve passing through the dotted circles).

\item[New Ordering Approach.]
The 90\% CL
confidence belt
is the region between the two dash-dot-dotted lines in Fig.~\ref{fig4}
and implies an upper limit of 2.6 events
for
$ \mu + b $
if $n_{\mathrm{obs}}=0$.
The corresponding exclusion curve is plotted in Fig.~\ref{fig5}
(the solid curve passing through the dotted triangles).

\end{description}

One can see that the four exclusion curves in Fig.~\ref{fig5}
obtained assuming an unknown background with the four different
statistical methods listed above
do not differ much.
This is a good sign and in practice tends to support the
exclusion curve obtained from
the standard confidence belt for upper confidence limits,
which coincides with the bayesian exclusion curve
plotted in Fig.~\ref{fig1}.

However,
it must be noted that
the choice to use the Unknown Background approach has been done on the basis of
the result of the experiment.
This procedure introduces some overcoverage,
as in the flip-flopping policy discussed by Feldman and Cousins~\cite{FC98}.

For example,
one can consider the possibility to take
$ b = n_{\mathrm{obs}} $
for
$ n_{\mathrm{obs}} < 3 $
and
$ b = 2.88 $
for
$ n_{\mathrm{obs}} \geq 3 $.
Using the standard confidence belts for upper confidence limits
for $b=0,1,2,2.88$ shown in Fig.\ref{fig6},
one obtain the confidence belt below the thick solid line in Fig.\ref{fig6}.
It is clear from the figure that this policy leads to overcoverage
for $ \mu \lesssim 3.3 $,
that is the interesting region in practice.
The value of the induced overcoverage is shown in Fig.\ref{fig7}
where I plotted the sum of the probabilities corresponding to the $n$'s
lying in the acceptance interval for a given $\mu$
of the confidence belt below the thick solid line in Fig.\ref{fig6}
as a function of $\mu$.
One can see that for $\mu \lesssim 3.3$
this sum is even bigger than one!
This is due to the fact that the probability
has not been properly normalized
to take into account that $b$ depends on $n$~\footnote{One
can also calculate the coverage of the confidence belt
below the thick solid line in Fig.\ref{fig6}
for different fixed values of $b$.
In this case one obtains
overcoverage for small values of $\mu$ if $b$ is large
(for example, $\mu=2$ and $b=2.88$)
and undercoverage for large values of $\mu$ if $b$ is small
(for example, $\mu=4$ and $b=0$).
However,
this calculation is in contradiction
with the assumption that $b$ is fixed by the procedure,
\textit{i.e.}
$b=n$ for $n<3$ and $b=2.88$ for $n\geq3$.}.
(The fact that for $\mu \gtrsim 3.3$ the sum of the probabilities
is bigger than 0.90 shows the unavoidable overcoverage
in the case of a Poisson processes due to the discreteness of $n$.
See, for example, Refs.\cite{PDG96,PDG98,FC98}.)

In conclusion of this Section,
we have seen that
the Unknown Background approach
is allowed and the corresponding exclusion curves
obtained with four different
statistical methods do not differ much.
In practice this result represents a support for the
bayesian exclusion curve
plotted in Fig.~\ref{fig1}.
However,
it is necessary to emphasize that the Unknown Background approach
is conservative for two reasons:
1) The upper confidence limit for the mean $\mu$ neutrino oscillation events
is taken to be equal to
the upper confidence limit for $ \mu + b $,
where $b$ is the unknown mean background events;
2) The choice to use the Unknown Background approach is done on the basis of
the result of the experiment,
introducing some overcoverage.
Hence, we can consider the exclusion curves
obtained with the Unknown Background approach
as \emph{ultra-conservative}.

\section{Sensitivity}
\label{Sensitivity}

In their Unified Approach paper~\cite{FC98}
Feldman and Cousins
suggested that in the cases in which the measurement is less than
the estimated mean background,
the experimental collaboration should report
also the sensitivity of the experiment,
defined as ``the average upper limit that would be obtained
by an ensemble of experiments with the expected background
and no true signal''.
This is also recommended by the Particle Data Group~\cite{PDG98}.

From the definition,
the sensitivity $\mu_s$ of an experiment
measuring a Poisson process with an expected mean background $b$
with negligible error
is given by\footnote{This formula can be easily generalized to the case
of an expected mean background $b=\overline{b}\pm\sigma_b$
with non-negligible error $\sigma_b$:
$
\mu_s(\overline{b},\sigma_b,\alpha)
=
\sum_{n=0}^{\infty}
\mu_1(n;\alpha)
\
P(n|\mu=0;\overline{b},\sigma_b)
$,
with
$P(n|\mu;\overline{b},\sigma_b)$
given by Eq.(\ref{pnmu1}).}
\begin{equation}
\mu_s(b,\alpha)
=
\sum_{n=0}^{\infty}
\mu_1(n;b,\alpha)
\
P(n|\mu=0,b)
\,,
\label{sensitivity}
\end{equation}
where $\alpha$ is the confidence level
of the upper limits $\mu_1(n;b,\alpha)$.
The sensitivity $\mu_s$
depends on the values of the upper limits $\mu_1(n;b,\alpha)$,
which are different in the different methods
for calculating the confidence belt
(\textit{e.g.}
in the Unified Approach
and
in the New Ordering Approach).

The sensitivity of the KARMEN~2 experiment for $\alpha=90\%$
is
$\mu_s=4.4$~\cite{KARMEN}
in the Unified Approach and
$\mu_s=4.7$~\cite{CG98}
in the New Ordering Approach.
The corresponding sensitivity curves are shown in Fig.~\ref{fig1}
(the solid curves passing through the empty circles and triangles, respectively).

One could think of considering the sensitivity curve of the KARMEN~2
experiment as its exclusion curve.
However, this is \emph{wrong}
for the following reasons.
From Eq.(\ref{sensitivity})
it is clear that
\emph{$\mu_s$
is the expected value of the upper limit $\mu_1$
under the hypothesis that $\mu=0$}.
This is the only correct meaning of $\mu_s$
Even though the value of $\mu_s$
depends on the confidence level $\alpha$,
the interval $[0,\mu_s]$
does not have a confidence level $\alpha$
from a frequentist point of view.

Notice that $\mu_s$ does not depend on any experimental result
and can be calculated for real as well as hypothetical experiments.
If the sensitivity curve corresponding to $\mu_s(b,\alpha)$
could be interpreted as an exclusion curve with confidence level $\alpha$,
one could obtain exclusion curves (\textit{i.e.} experimental results)
without performing any experiment!
This apparent paradox is solved by noting that
$\mu_s$
is \emph{not} calculated by assuming that the number of observed events is zero
(all values of $n$ are taken into account in Eq.(\ref{sensitivity}))
but assuming that the true value of $\mu$ is zero
and if the true value of $\mu$
would be known
it would be useless to do any experiment to measure it.

Each sensitivity curve in Fig.~\ref{fig1}
is shifted to the right with respect to the corresponding exclusion curve,
although it is calculated under the assumption that $\mu=0$,
because
the sensitivity curve takes into account the possibility that an experiment
with mean background $b$
can measure a number of events $n>b$,
even if $\mu=0$.
For example,
in the Unified Approach with $b=2.88$
one has 
$ \mu_1(n=3;\alpha=0.90) = 4.5 $,
$ \mu_1(n=4;\alpha=0.90) = 5.7 $,
$ \mu_1(n=5;\alpha=0.90) = 7.1 $,
and so on.
On the other hand,
since
$ \mu_1(n=2;\alpha=0.90) = 3.1 $,
an experiment measuring two events,
in good agreement with the mean expected background $b=2.88$,
can draw an exclusion curve that is significantly
more stringent than the sensitivity curve for $b=2.88$.

Since the sensitivity curve of a neutrino oscillation experiment
can be calculated before doing the experiment,
it is useful in order to plan future experiments
that cover approximately the region of interest
in the plane of the neutrino mixing parameters
$\sin^22\theta$ and $\Delta{m}^2$.

\section{Conclusions}
\label{Conclusions}

The statistical interpretation
of the null result of the KARMEN~2 neutrino oscillation experiment
is rather problematic because
no events were observed with a mean expected background of
$ 2.88 \pm 0.13 $ events~\cite{KARMEN}.
The exclusion curves obtained with the Bayesian Approach
and with the Unified Approach~\cite{FC98}
are significantly different and yield contradicting
indications on the compatibility of the KARMEN~2 result
with the neutrino oscillation signal measured
in the LSND experiment~\cite{LSND}
(see Fig.~\ref{fig1}).
The analysis of the KARMEN~2 null result with the New Ordering Approach~\cite{CG98},
which is a frequentist method with correct coverage as the Unified Approach,
yields an exclusion curve close to the one obtained with the
Bayesian Approach.
In this way,
the undesirable discrepancy between frequentist and bayesian
interpretations of the KARMEN~2 null result
is removed.
However,
some concern still persist on the
interpretation of the fact that none of the expected background events
have been observed in the KARMEN~2 experiment.
This concern will increase in the future if the
KARMEN~2 experiment will continue to observe significantly less
background events than expected.
Hence, it is interesting to investigate other
possibilities for the statistical interpretation
of the KARMEN~2 null result.

In this paper
I have discussed three alternative statistical interpretation
of the KARMEN~2 null result:
the Larger Background Error possibility in Section~\ref{Larger Background Error},
the Unknown Background possibility in Section~\ref{Unknown Background}
and the Sensitivity possibility in Section~\ref{Sensitivity}.

In Section~\ref{Larger Background Error}
I have presented the formalism that allows to take into account
the error of the expected mean background.
However,
the conclusion of this section is that taking into account
the error of the expected mean background
in the KARMEN~2 experiment,
even if it is wrong by an order of magnitude,
does not help in solving the problem of the statistical interpretation
of the result of this experiment,
because
the resulting exclusion curves are practically equivalent
to the ones obtained assuming no error
for the expected mean background.

The Unknown Background approach discussed in Section~\ref{Unknown Background}
gives ultra-conservative exclusion curves,
which in the case of the KARMEN~2 experiment
tend to support the exclusion curve obtained with the Bayesian Approach.
Obtaining ultra-conservative exclusion curves
is generally not desirable,
but could be a safe choice in controversial cases as that of the KARMEN~2 experiment.

The possibility to consider a Sensitivity curve as an exclusion curve
is discussed in Section~\ref{Sensitivity}
and is shown to be wrong.
Since the sensitivity curve of a neutrino oscillation experiment
can be calculated before doing the experiment,
its usefulness lies in the possibility
to plan future experiments
in order to cover approximately the region of interest
in the plane of the neutrino mixing parameters
$\sin^22\theta$ and $\Delta{m}^2$.

Finally,
I would like to remark that
the direct comparison of exclusion curves and allowed regions
obtained with different statistical methods
does not have a precise statistical significance.
Hence,
such a comparison cannot be used to combine the results of
different experiments
or to infer with some known confidence level
a contradiction between the results of different experiments
when the comparison is done on the border of the
exclusion curves and of the allowed regions.
Hence,
the comparison of the
KARMEN~2 exclusion curves
and
the LSND-allowed region,
which was obtained with a different statistical analysis
(see Ref.~\cite{LSND}),
must be done with great caution.
In the future,
if the KARMEN~2 experiment will continue to observe no neutrino oscillations,
it will be possible to claim a contradiction between the results of
the two experiments only when the exclusion curves obtained from the
result of the KARMEN~2 experiment with different statistical methods
will produce similar results and will lie well on the left of
the region allowed by the results of the LSND experiment.

\begin{figure}[h]
\refstepcounter{figure}
\label{fig1}
Fig.~\ref{fig1}.
90\% CL
exclusion curves
in the plane of the neutrino oscillation parameters
$\sin^22\theta$--$\Delta{m}^2$
corresponding to the
null result of the KARMEN~2 experiment~\cite{KARMEN}.
The solid curves passing through the filled squares, circles and triangles
are obtained with the Bayesian Approach, the Unified Approach
and the New Ordering Approach, respectively.
The solid curves passing through the empty circles and triangles
are the sensitivity curves obtained with the Unified Approach
and the New Ordering Approach, respectively.
The shadowed area
is the region allowed at 90\% CL
by the results of the LSND experiment \cite{LSND}
and the dashed, dash-dotted and dash-dot-dotted curves are the
90\% CL
exclusion curves of the
Bugey \cite{Bugey}, BNL E776 \cite{BNL E776} and CCFR \cite{CCFR} experiments,
respectively.
\end{figure}

\begin{figure}[h]
\refstepcounter{figure}
\label{fig2}
Fig.~\ref{fig2}.
Confidence belts for 90\% CL
obtained with the Unified Approach for
$\overline{b}=2.88$
and
$\sigma_b=0$
(the region between the two solid lines)
and
$\sigma_b=1.3$
(the region between the two dashed lines).
\end{figure}

\begin{figure}[h]
\refstepcounter{figure}
\label{fig3}
Fig.~\ref{fig3}.
Confidence belts for 90\% CL
obtained with the New Ordering Approach for
$\overline{b}=2.88$
and
$\sigma_b=0$
(the region between the two solid lines)
and
$\sigma_b=1.3$
(the region between the two dashed lines).
\end{figure}

\begin{figure}[h]
\refstepcounter{figure}
\label{fig4}
Fig.~\ref{fig4}.
Confidence belts for 90\% CL
obtained in the case of unknown background
(see Section~\ref{Unknown Background})
with the standard method for upper confidence limits
(the region under the solid line),
with the standard method for central confidence intervals
(the region between the two dashed lines),
with the Unified Approach
(the region between the two dash-dotted lines)
and
with the New Ordering Approach
(the region between the two dash-dot-dotted lines).
The lower confidence limits $ (\mu + b)_2 $
calculated with the Unified Approach and the New Ordering Approach
coincide for $n<3$.
All the confidence belts include the origin of the coordinates
$n$, $\mu+b$.
\end{figure}

\begin{figure}[h]
\refstepcounter{figure}
\label{fig5}
Fig.~\ref{fig5}.
90\% CL
exclusion curves
obtained in the case of unknown background
(see Section~\ref{Unknown Background})
with the standard method for upper confidence limits
(the solid curve passing through the filled squares),
with the standard method for central confidence intervals
(the solid curve passing through the dotted squares),
with the Unified Approach
(the solid curve passing through the dotted circles)
and
with the New Ordering Approach
(the solid curve passing through the dotted triangles).
\end{figure}

\begin{figure}[h]
\refstepcounter{figure}
\label{fig6}
Fig.~\ref{fig6}.
Standard 90\% CL
confidence belts for upper confidence limits
for $b=0,1,2,2.88$
(the region below the
dashed, dotted, dash-dotted and dash-dot-dotted line, respectively).
The region below the thick solid line
is the confidence belt obtained taking
$ b = n $
for
$ n < 3 $
and
$ b = 2.88 $
for
$ n \geq 3 $.
\end{figure}

\begin{figure}[h]
\refstepcounter{figure}
\label{fig7}
Fig.~\ref{fig7}.
Sum of the probabilities corresponding the $n$'s
lying in the acceptance interval for a given $\mu$
of the confidence belt below the thick solid line in Fig.\ref{fig6}
as a function of $\mu$.
\end{figure}

\newpage

\begin{minipage}[p]{0.95\linewidth}
\begin{center}
\mbox{\epsfig{file=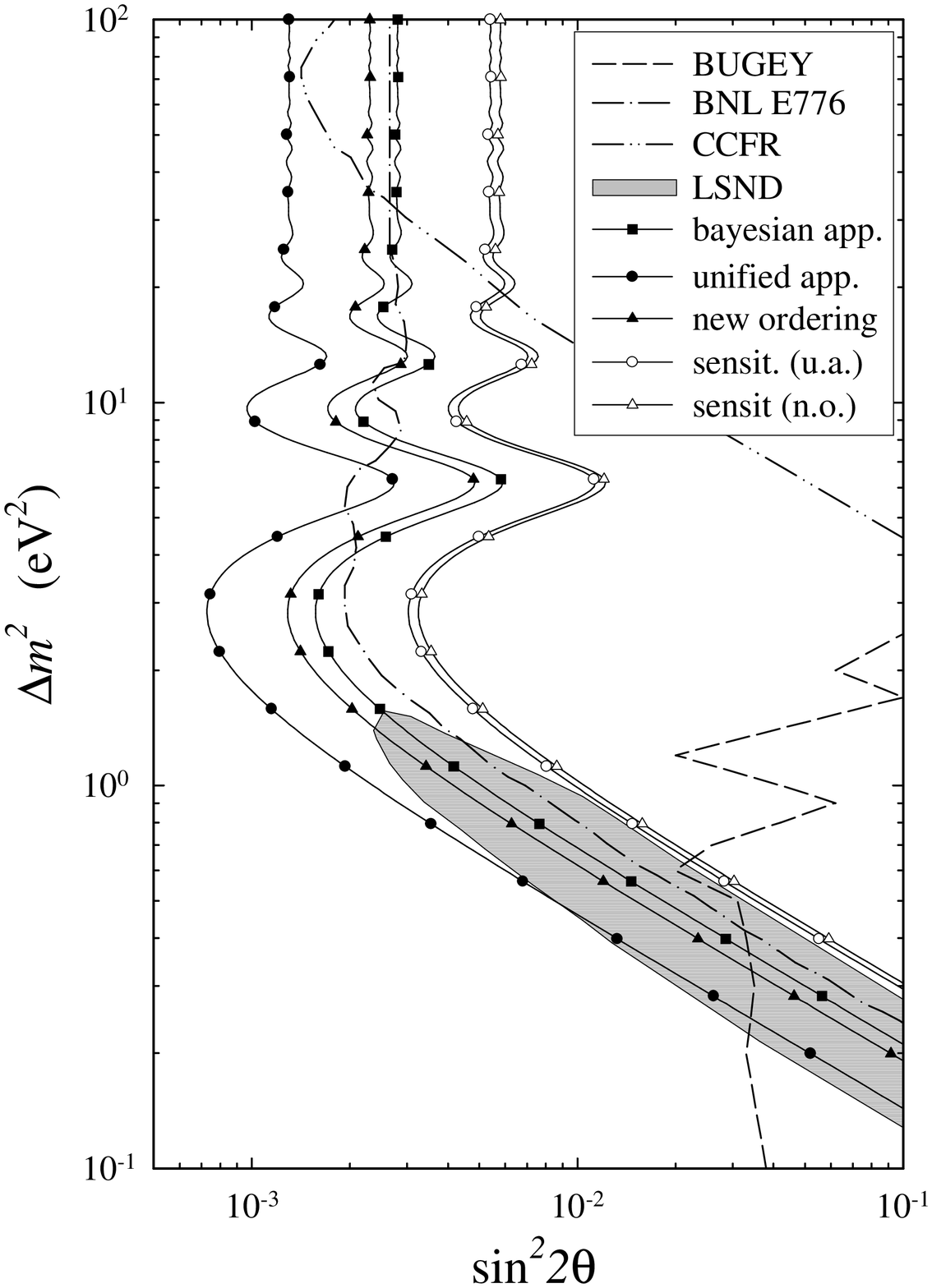,width=0.95\linewidth}}
\end{center}
\end{minipage}
\begin{center}
\Large Figure \ref{fig1}
\end{center}

\begin{minipage}[p]{0.95\linewidth}
\begin{center}
\mbox{\epsfig{file=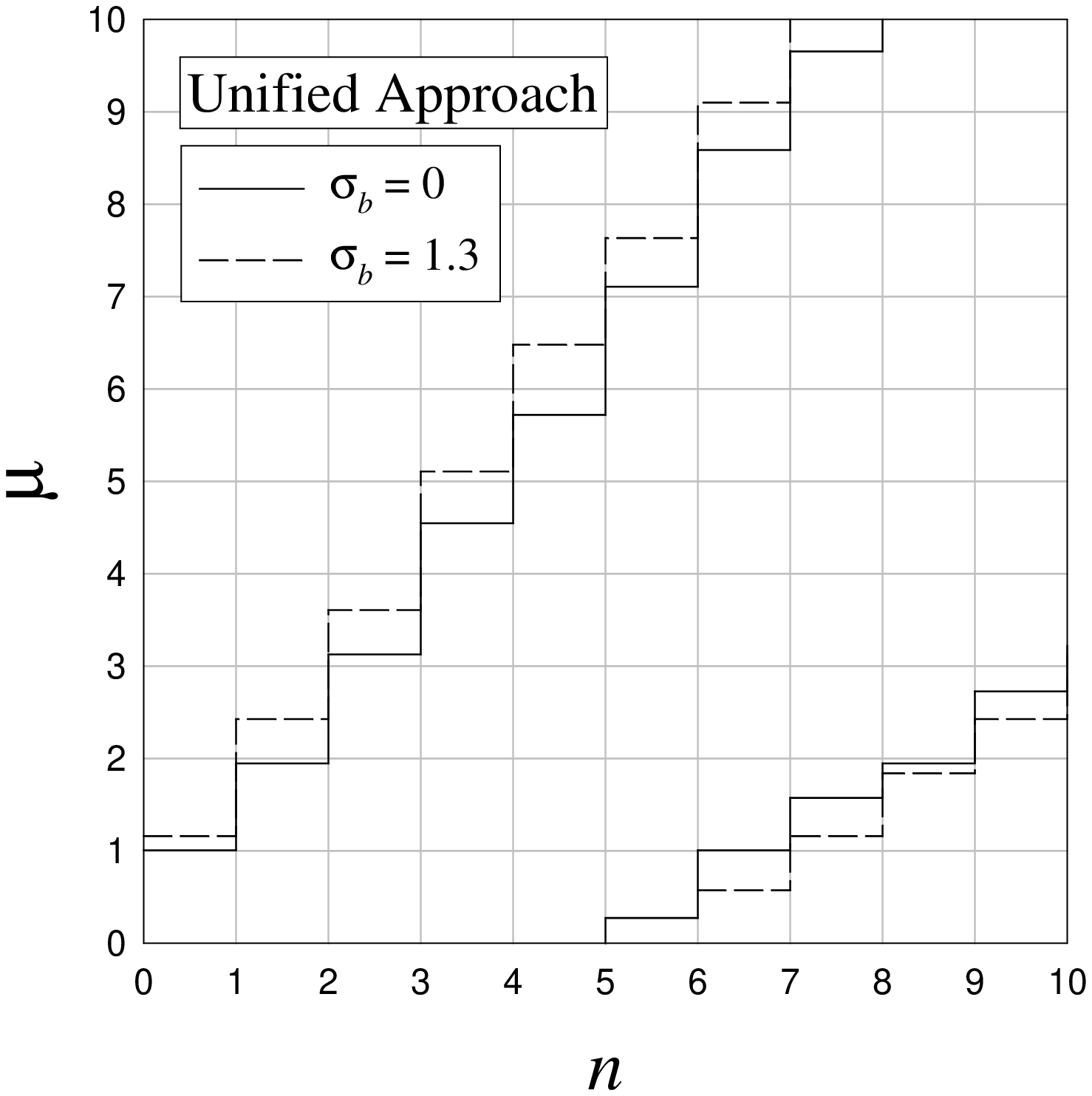,width=0.95\linewidth}}
\end{center}
\end{minipage}
\begin{center}
\Large Figure \ref{fig2}
\end{center}

\begin{minipage}[p]{0.95\linewidth}
\begin{center}
\mbox{\epsfig{file=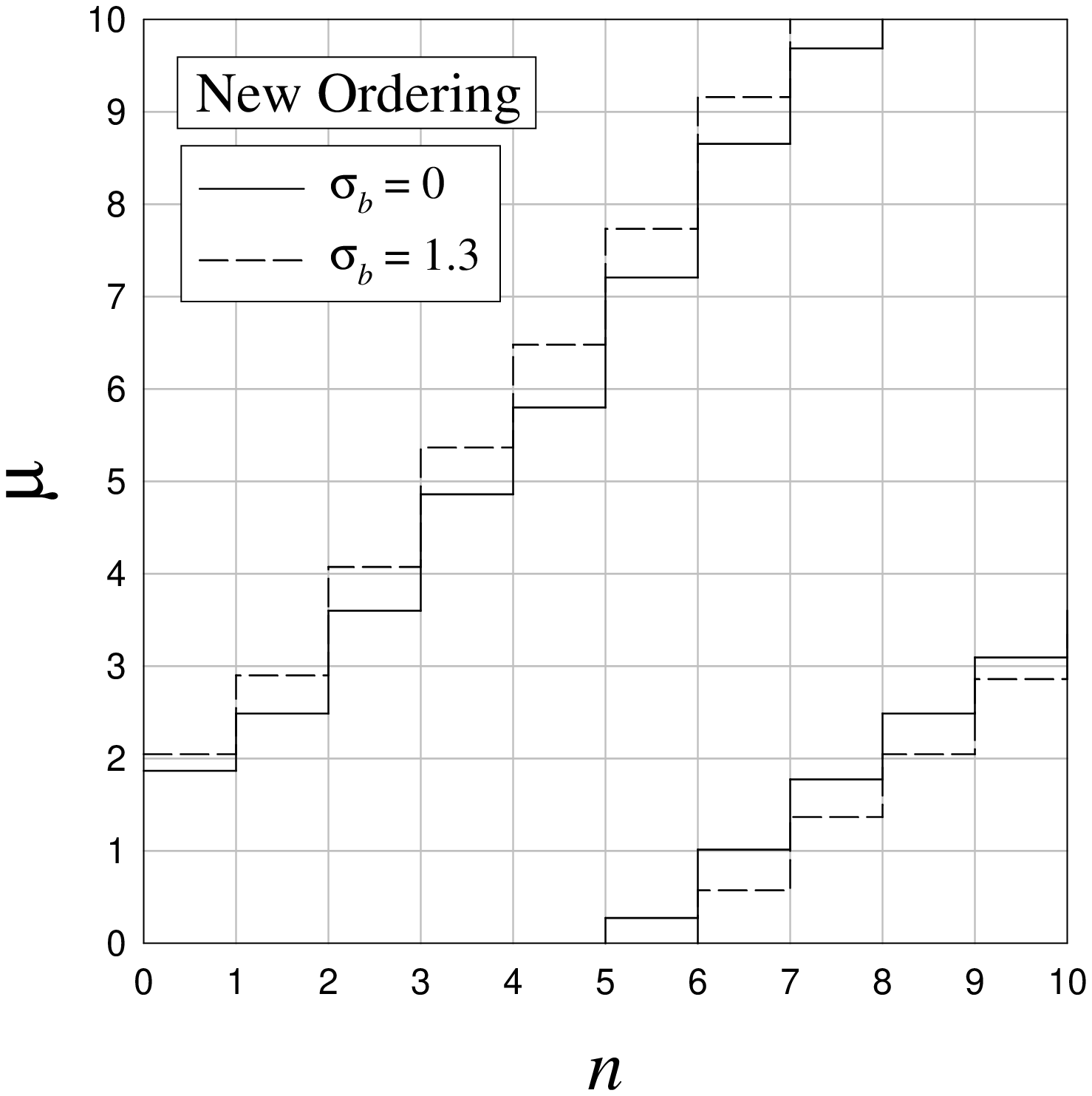,width=0.95\linewidth}}
\end{center}
\end{minipage}
\begin{center}
\Large Figure \ref{fig3}
\end{center}

\begin{minipage}[p]{0.95\linewidth}
\begin{center}
\mbox{\epsfig{file=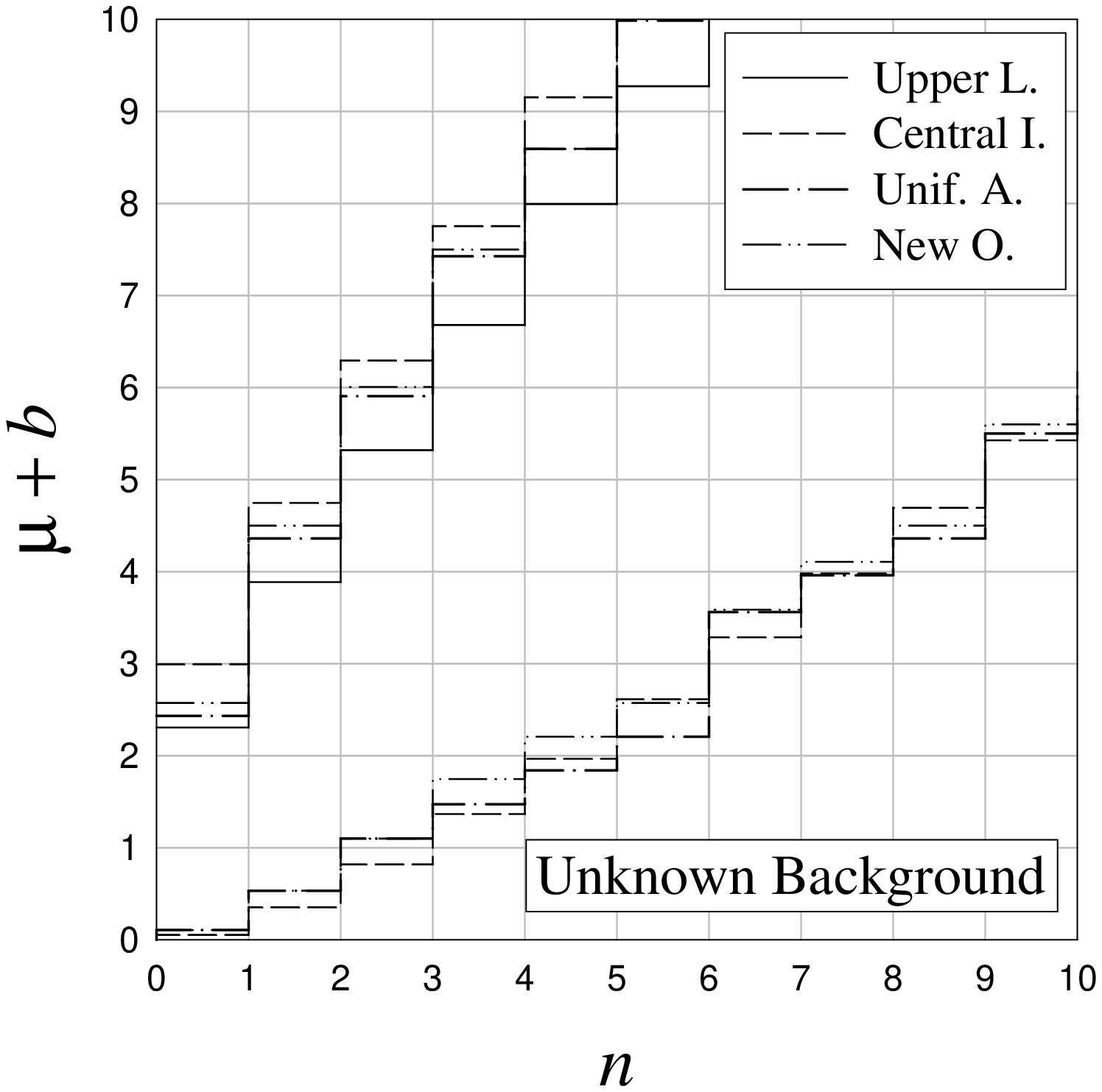,width=0.95\linewidth}}
\end{center}
\end{minipage}
\begin{center}
\Large Figure \ref{fig4}
\end{center}

\begin{minipage}[p]{0.95\linewidth}
\begin{center}
\mbox{\epsfig{file=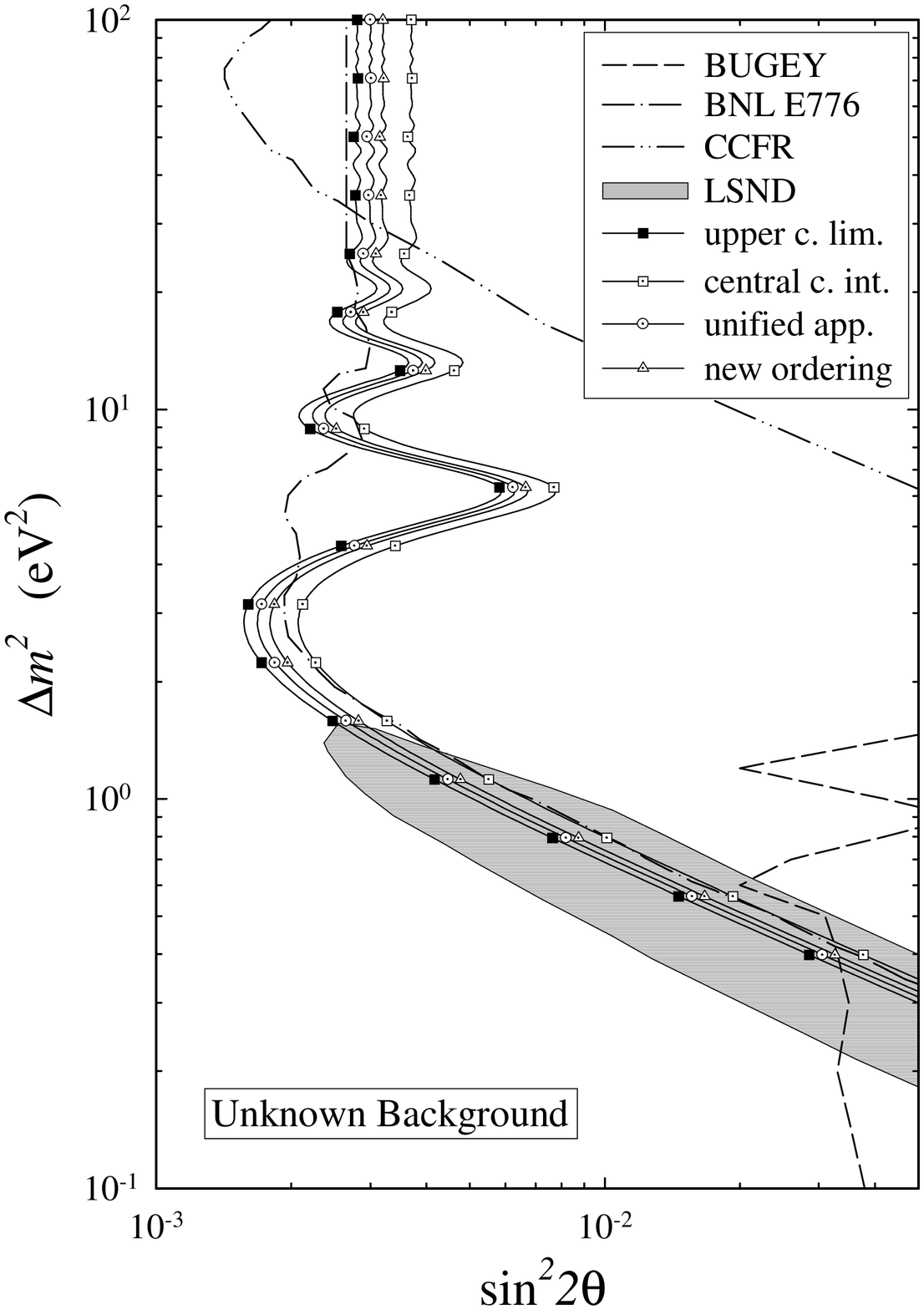,width=0.95\linewidth}}
\end{center}
\end{minipage}
\begin{center}
\Large Figure \ref{fig5}
\end{center}

\begin{minipage}[p]{0.95\linewidth}
\begin{center}
\mbox{\epsfig{file=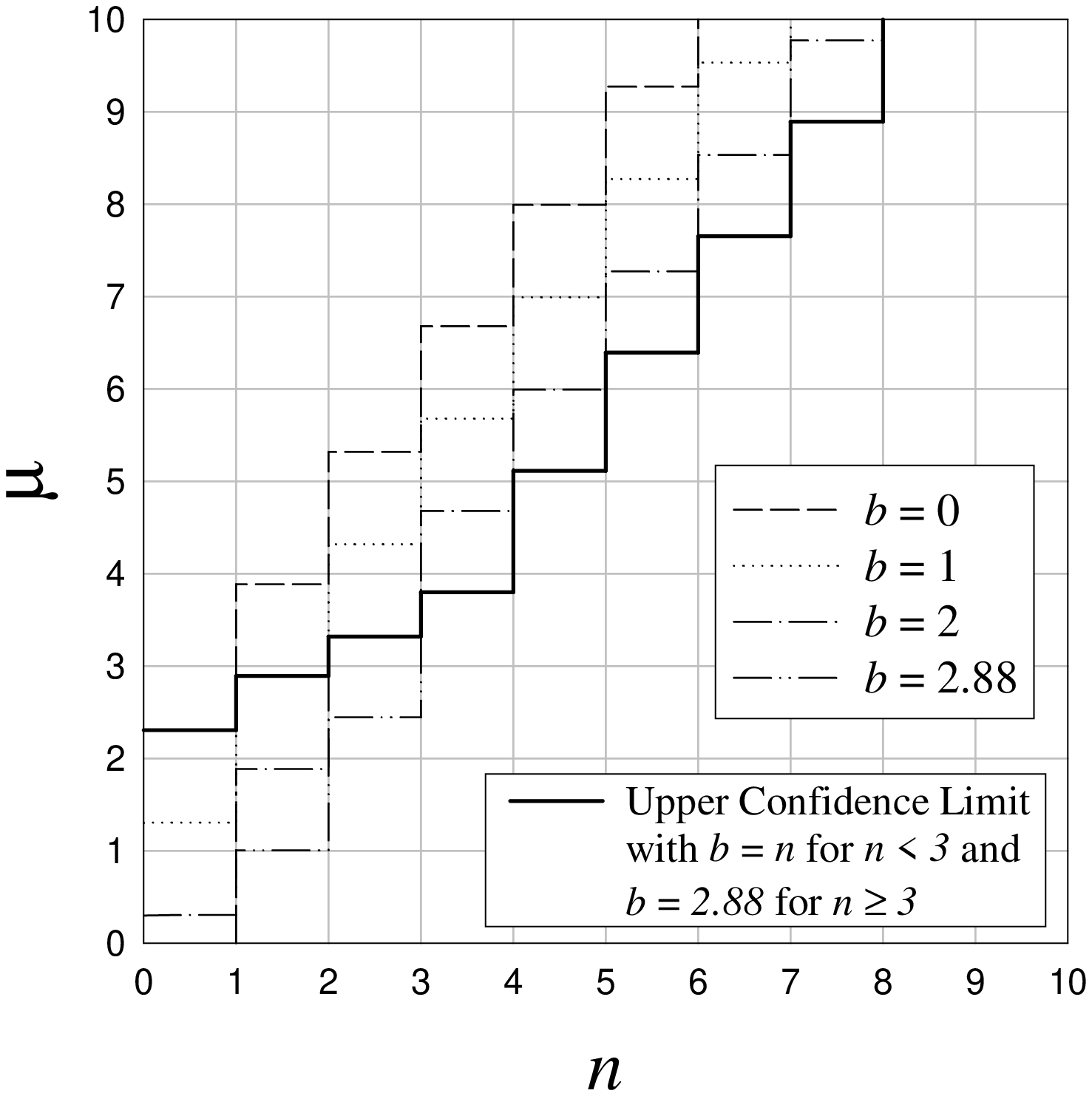,width=0.95\linewidth}}
\end{center}
\end{minipage}
\begin{center}
\Large Figure \ref{fig6}
\end{center}

\begin{minipage}[p]{0.95\linewidth}
\begin{center}
\mbox{\epsfig{file=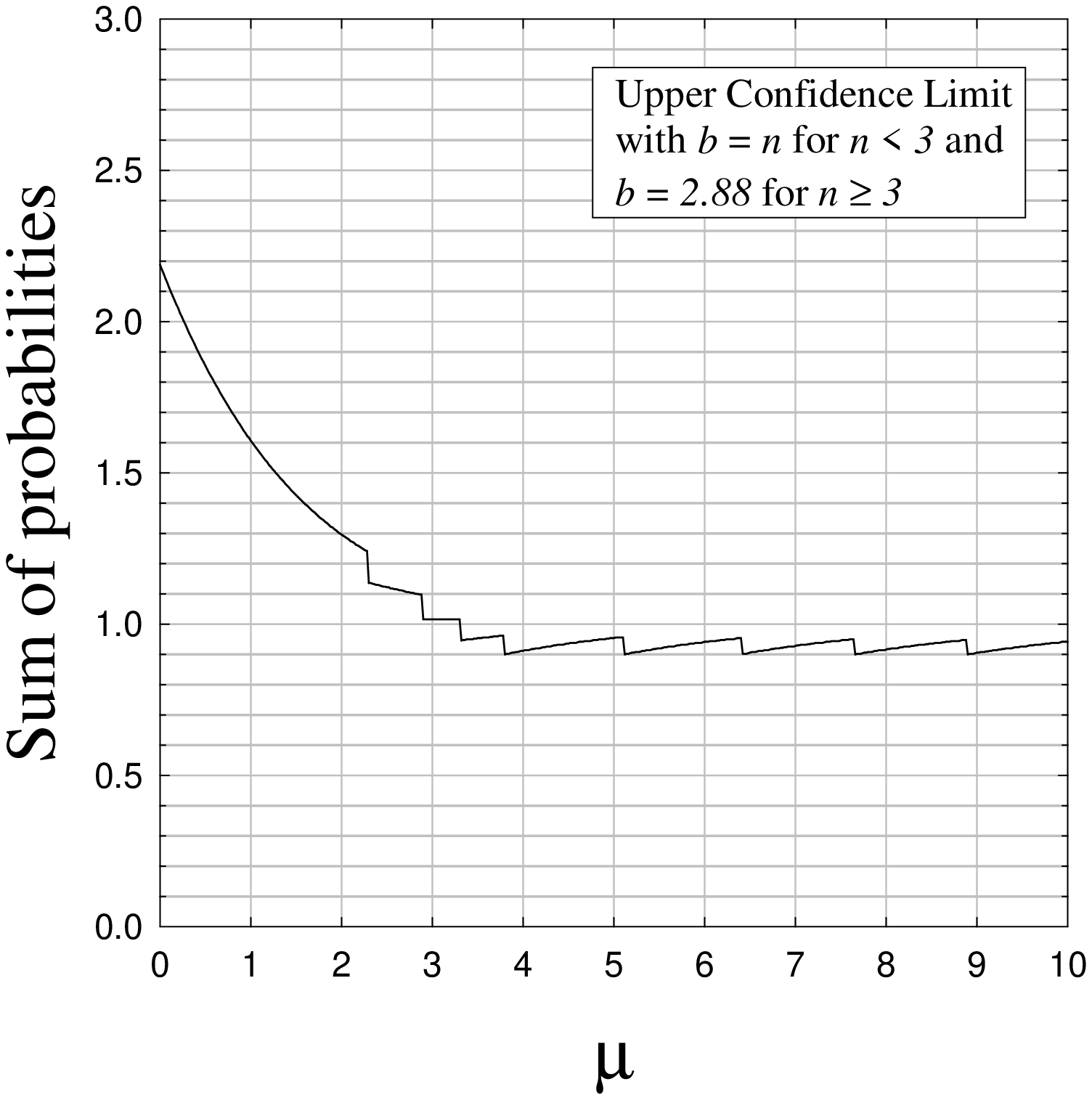,width=0.95\linewidth}}
\end{center}
\end{minipage}
\begin{center}
\Large Figure \ref{fig7}
\end{center}

\end{document}